\documentclass[12pt,twoside]{article}
\usepackage{rotating,booktabs}
\usepackage{amsmath}
\usepackage{graphicx}
\usepackage{dcolumn}
\usepackage{bm}
\usepackage{amssymb}
\usepackage{latexsym}
\def\Journal#1#2#3#4{{#1} {\bf #2}, #3 (#4)}  

\def\NPA{{ Nucl. Phys.} {\bf A}}

\def\PLB{{ Phys. Lett.} {\bf B}}

\def\PRC{{ Phys. Rev.} { C}}

\def\PRT{ Phys. Rep.}
\def\JPG{{ J. Phys.} G}

\def\J{J/\psi}

\newcommand{\be}{\begin{equation}}
\newcommand{\ee}{\end{equation}}
\newcommand{\bea}{\begin{eqnarray}}
\newcommand{\eea}{\end{eqnarray}}
\newcommand{\bean}{\begin{eqnarray*}}
\newcommand{\eean}{\end{eqnarray*}}

\begin{document}



\title{Subthreshold $\rho$ contribution in $J/\psi$ decay to
 $\omega\pi\pi$  and $K\bar{K}\pi$
                }

\vspace{2cm}
\author{F. Q. Wu$^{a,b,d}$ and B. S. Zou$^{a,b,c}$\\
a) CCAST (World Laboratory), P. O. Box 8730, Beijing 100080,China
\footnote{mailing address; E-mail: wufq@ihep.ac.cn, zoubs@ihep.ac.cn}\\
b) Institute of High Energy Physics, CAS, Beijing 100049, China\\
c) Institute of Theoretical Physics, CAS, Beijing 100080, China\\
d) Graduate School, Chinese Academy of Sciences, Beijing
100049,China}

\date{\today}


\maketitle

\begin{abstract}
We carry out a theoretical and Monte Carlo study on the $J/\psi$
decays into $\omega\pi\pi$ and  $K\bar{K}\pi$ through intermediate
subthreshold  $\rho$  meson by using SU(3)-symmetric Lagrangian
approach.   It is found that the subthreshold $\rho$ contribution
is not negligible  and may have significant influence on partial
wave analysis of resonances in these channels, especially near the
$\omega \pi$ and $K \bar{K}$ thresholds.

\end{abstract}
\vspace{1cm}


\section{Introduction}
In recent years, much effort has been devoted to the study of
meson and baryon spectra. Of particular interests, the $J/\psi$
decays now provide an excellent source of information for studying
light hadron spectroscopy and searching for glueballs, hybrids,
and exotic states \cite{bes-review}. Several interesting
near-threshold structures were observed and studied, such as the
broad $\sigma$ near $\pi\pi$ threshold \cite{bes-sigma}, the broad
$\kappa$ near $K\pi$ threshold \cite{bes-kappa}, the narrow
$f_0(980)$ peak near $\bar KK$ threshold \cite{bes-f980}, and the
narrow structure near $\bar pp$ threshold \cite{bes-p-anti-p}.
While various mechanisms were proposed to explain these structures
\cite{sigma1}, the conventional t-channel meson exchange final
state interaction mechanism \cite{rho-exchange,Wu:2003wf} can give
consistent explanation for all these structures. The t-channel
$\rho$ meson exchange was found to play a very important role for
all these structures.

In this paper, we want to address two other puzzling
near-threshold phenomena in $J/\psi$ decays.  The first one is the
``$b_1$ puzzle" in the $J/\psi \to \omega \pi \pi$. Both DM2
Collaboration \cite{dm2 5pi} and BESII Collaboration
\cite{bes-sigma} obtained a much broader width for the
near-$\omega\pi$-threshold resonance $b_1(1235)$: while PDG gives
the width $(142\pm 9)$ MeV \cite{PDG}, the DM2 and BESII gave it
as $(210\pm 19)$ MeV and $(195\pm 20)$ MeV, respectively. The
second puzzle is that there is a clear near-$\bar KK$-threshold
enhancement in both $J/\psi\to K^+K^-\pi^0$ and $J/\psi\to
K_SK^\pm \pi^\mp$ Dalitz plots from DM2 data \cite{dm2-KKpi}
although the structure was not addressed in the paper possibly due
to uncertainty of background contribution. From conservation laws
for strong interaction, the $\bar KK$ here should have isospin 1
and spin-parity $1^-$. There is no known resonance of these
quantum numbers very close to $\bar KK$ threshold.

\begin{figure}[htbp] 
\begin{center}
\includegraphics[scale=0.5]{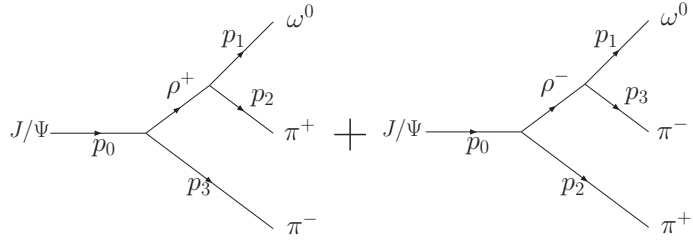}
\caption{Diagrams for $J/\psi \rightarrow \omega \pi^+ \pi^-$
decay with subthreshold $\rho$ exchange. }
 \label{fig:omega}
\end{center}
\end{figure}

\begin{figure}[htbp] \vspace{-1.cm}
\begin{center}
\includegraphics[scale=0.5]{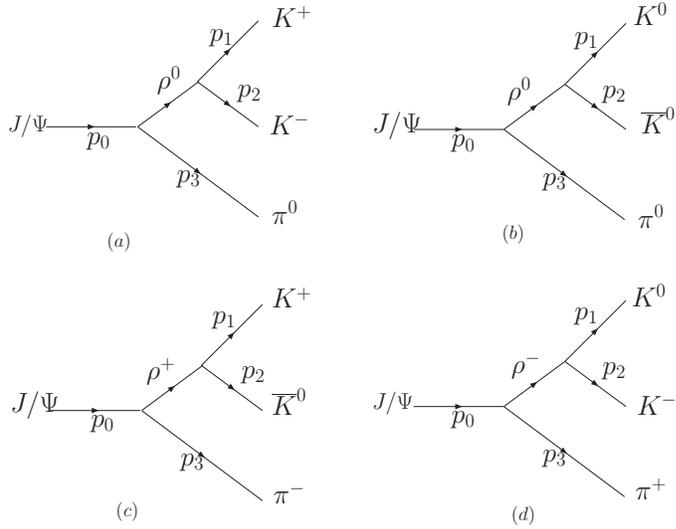}
\caption{Diagrams for $J/\psi \rightarrow  K \bar{K} \pi$
decay with subthreshold $\rho$ exchange. }
 \label{fig:k}
\end{center}
\end{figure}

It is rather tempting to claim some new near-threshold resonances
here. But before claiming any new physics from the seemingly
puzzling phenomena, one should investigate all possible
conventional mechanisms to see if the phenomena can be interpreted
within the existing theoretical framework. Motivated by this idea,
here we investigate the subthreshold $\rho$ contribution to these
channels through diagrams shown in Fig.\ref{fig:omega} and
Fig.\ref{fig:k} for $J/\psi \to \omega \pi \pi$ and $J/\psi\to
\bar KK\pi$, respectively. There are good reasons for considering
this mechanism. The $J/\psi\to\rho\pi$ decay has the largest
branching ratio among the known two-body decay channels of the
$J/\psi$ \cite{PDG}. The $\rho KK$ and $\rho\omega\pi$ couplings
are well determined to be large.

This paper is organized as follows. In section II, we present the
method and formulation for calculation and Monte Carlo simulation.
The numerical results and discussions are given in section III.

\section{Method and formulation}

The Feynman diagrams for relevant processes with the subthreshold
$\rho$ contributions  are depicted in Fig.\ref{fig:omega} ($J/\psi
\to \omega\pi^+\pi^-$ ) and Fig.\ref{fig:k} ($J/\psi \to
K\bar{K}\pi$ ). For the vector-vector-pseudoscalar (VVP) and the
pseudoscalar-pseudoscalar-vector (PPV) couplings, we use the
SU(3)-symmetric Lagrangians as in \cite{Wu:2003wf,lagrangian}
 \bea
  \label{eq:LVVP}
{\cal L}_{VVP} &=& \frac{G}{\sqrt{2}}\epsilon^{\mu \nu \alpha
\beta}\langle
\partial_{\mu} V_{\nu} \partial_{\alpha} V_{\beta} P \rangle,
\\
 \mathcal{L}_{PPV} &=&-\frac{1}{2}iG^{\prime} \langle  [P,\partial_\mu
P]V^\mu \rangle,
 \eea
  where $\langle \ldots\rangle$ means $SU(3)$ trace, $G$ and $G^\prime$ are the
coupling constants,  and $P$ is the $3 \times 3$ matrix representation
of the pseudoscalar meson octet, here $P = \lambda_a P^a$, $a = 1, . .... .
, 8$ and $\lambda_a$ are the $3 \times 3$ generators of SU(3). A
similar definition of $V_\nu$ is used for the vector meson octet.

In the Gell-Mann representation, the relevant effective Lagrangians
are
 \bea
      \mathcal{L}_{\psi \rho \pi} &=& g_{\psi\rho\pi}\epsilon^{\mu \nu \alpha
\beta}p^\mu_\psi e^\nu(\psi) p^\alpha_\rho e^\beta(\rho), \label{psi rho pi}\\
      \mathcal{L}_{\omega \rho \pi} &=& g_{\omega\rho\pi}\epsilon^{\mu \nu \alpha
\beta}p^\mu_\omega e^\nu(\omega) p^\alpha_\rho e^\beta(\rho), \label{omega rho pi}\\
%
\mathcal{L}_{\rho\pi\pi} &=& g_{\rho\pi\pi} [(p_{\pi^+}^\mu
-p_{\pi^-}^\mu )\rho^0_{\mu} + (p_{\pi^-}^\mu -p_{\pi^0}^\mu
)\rho^+_{\mu} +
(p_{\pi^0}^\mu -p_{\pi^+}^\mu )\rho^-_{\mu}] \\
  \mathcal{L}_{\rho K \bar{K}} &=& g_{\rho K
\bar{K}} [(p_{K^+}^\mu - p_{K^-}^\mu)
  \rho^0_\mu + (p_{\bar{K}^0}^\mu - p_{{K}^0}^\mu) \rho^0_\mu
] + \nonumber \\ && \sqrt{2}~ g_{\rho K \bar{K}} (p_{K^0}^\mu
- p_{K^-}^\mu) \rho^+_\mu + \sqrt{2}~ g_{\rho K \bar{K}}
(p_{K^+}^\mu - p_{\bar{K}^0}^\mu) \rho^-_\mu  \label{psi rho k
k}
 \eea
 where $g_{\rho K
\bar{K}} = \frac{1}{2} g_{\rho\pi\pi}= G^\prime$ due to flavor SU(3) symmetry.
 Using these Lagrangians,  we are able to construct following amplitudes $T_1$,
$T_{2a}$, $T_{2b}$, $T_{2c}$ and $T_{2d}$ corresponding the
diagrams in Fig.\ref{fig:omega}, Fig.\ref{fig:k}a,
Fig.\ref{fig:k}b, Fig.\ref{fig:k}c and Fig.\ref{fig:k}d,
respectively:
  \bea  T_1 &=& -g_{\psi \rho \pi} g_{\omega \rho \pi}\{
  \epsilon_{\mu \nu \alpha \beta} p_0^\mu e^{*\nu}(\psi) (p_1+p_2)^\alpha
  \epsilon^{\gamma\beta\lambda \sigma} (p_1+p_2)_\gamma p_{1\lambda} e_{\sigma}(\omega) \nonumber
  \\ && \frac{1}{(p_1+p_2)^2-m_\rho^2 + i m_\rho \Gamma_\rho}+
   \frac{1}{(p_1+p_3)^2-m_\rho^2 + i m_\rho \Gamma_\rho} \nonumber
  \\ &&
  \epsilon_{\mu \nu \alpha \beta} p_0^\mu e^{*\nu}(\psi) (p_1+p_3)^\alpha
  \epsilon^{\gamma\beta\lambda \sigma} (p_1+p_3)_\gamma p_{1\lambda} e_{\sigma}(\omega)
   \} F_{\psi \rho \pi}F_{\omega \rho \pi} , \label{T1}\\
  T_{2a} &=&  T_{2b} = \frac{1}{\sqrt{2}} T_{2c} = \frac{1}{\sqrt{2}}T_{2d}
 \nonumber \\
 &=& -\frac{ 2g_{\psi \rho \pi} g_{\rho K \bar{K}}}{(p_1+p_2)^2-m_\rho^2 + i m_\rho \Gamma_\rho}
 \epsilon_{\mu \nu \alpha \beta} p_0^\mu e^{*\nu}(\psi)
 p_1^\alpha p_2^\beta F_{\psi \rho \pi}F_{\rho K
 \bar{K}}. \label{T2}
\eea
Here $\Gamma_\rho$ represents the width of $\rho$; $F_{\psi \rho
\pi}$, $F_{\omega \rho \pi}$  and $F_{\rho K \bar{K}}$ are
the form factors for the $\psi \rho \pi$, $\omega \rho \pi$ and
$\rho K \bar{K}$ vertices, respectively.

Usually, hadronic form factors should be applied to the
meson-meson-meson vertices because of the inner quark-gluon
structure of hadrons. It is well known that form factors play an
important role in many physics processes, such as $\pi\pi$
scattering \cite{Wu:2003wf},  NN interactions \cite{Machleidt},
$\pi N$ scattering \cite{Gross}, meson photo-production
\cite{Q.Zhao} etc.. Due to the difficulties in dealing with
nonperturbative QCD hadron structure, the form factors are
commonly adopted phenomenologically.

The most commonly used form factors for meson-meson-meson vertices
in $J/\psi$ decays are Blatt-Weisskopf barrier factors
$B_l(Q_{abc})$ \cite{VonHippel:1972fg,chung:93 98} for a decay
process $a\to b+c$ with orbital angular momentum $l$ between $b$
and $c$ mesons:
 \bea B_1(Q_{abc},R)\!&=&\!
\sqrt{Q^2_0\over Q_{abc}^2+Q^2_0} ~~~~\text{for} ~l=1,  \eea where
\begin{equation}
Q_{abc}^2=\frac{(s_{a}+s_{b}-s_{c})^{2}}{4s_{a}}-s_{b}
\end{equation}
is the magnitude of $\bf p_b$ or $\bf p_c$ in the rest system of
$a$; $s_i=E_i^2-{\bf p}^2_i$ with $E_i$ and ${\bf p}_i$ the energy
and the tree-momentum component of $p_i$, respectively. Here $Q_0$
is a hadron ``scale" parameter, $Q_0 = 0.197321/R$ GeV/c with $R$
reflecting the radius of the centrifugal barrier in fm. We take
$R=0.5$ fm for $\psi \rho \pi$ vertex and $R=0.5$ or 0.8 fm for
$\omega \rho \pi$ and $\rho \pi \pi$ vertices. For the sake of
convenience, we use the following shorthand notation for the
$F_{\psi \rho \pi}F_{\rho \text{m} \text{m} }$ in
Eqs.(\ref{T1},\ref{T2}) for various channels:
 \bea F_1 &\equiv& \left\{ \begin{array}{clc}
 B_1(R_{\psi \rho \pi}=0.5) B_1(R_{\rho
 \pi\pi}=0.5)&, & J/\psi \stackrel{\rho}{\to} \pi \pi \pi \\
 B_1(R_{\psi \rho \pi}=0.5) B_1(R_{\omega \rho \pi}=0.5)&, & J/\psi \stackrel{\rho}{\to} \omega \pi \pi
 \\
 B_1(R_{\psi \rho \pi}=0.5) B_1(R_{\rho K \bar{K}}=0.5)&, & J/\psi \stackrel{\rho}{\to} K \bar{K}\pi
 \end{array} \right.
 \eea
\bea F_2 &\equiv& \left\{ \begin{array}{clc}
 B_1(R_{\psi \rho \pi}=0.5) B_1(R_{\rho
 \pi\pi}=0.8)&, & J/\psi \stackrel{\rho}{\to} \pi \pi \pi \\
 B_1(R_{\psi \rho \pi}=0.5) B_1(R_{\omega \rho \pi}=0.8)&, & J/\psi \stackrel{\rho}{\to} \omega \pi \pi
 \\
 B_1(R_{\psi \rho \pi}=0.5) B_1(R_{\rho K \bar{K}}=0.8)&, & J/\psi \stackrel{\rho}{\to} K \bar{K}\pi
 \end{array} \right.
 \eea

The monopole form factor is also a frequently used s-channel form
factor \cite{Wu:2003wf,Liu,Titov}:
\begin{equation}
F(\Lambda, q)=\frac{\Lambda^2+m^2}{\Lambda^2+q^2},
\end{equation}
where $m$ and $q$ are the mass and the four-momentum of the
intermediate particle, respectively, and $\Lambda$ is the
so-called cut-off momentum that can be determined by fitting the
experimental data. We take a commonly used value $\Lambda$=1.5 GeV
for $\rho \pi \pi$ \cite{Wu:2003wf} and $\omega \rho \pi$
\cite{Titov} vertices. For $\rho K \bar{K}$ vertex,
$\Lambda$=1.5-4.5 GeV was used in literature
\cite{rho-exchange,Wu:2003wf}. It is possible that mesons
involving heavier quarks have smaller size and corresponding need
larger cut-off $\Lambda$ parameter for the relevant vertices. We
take $\Lambda$=1.5-4.5 GeV for $\J\rho\pi$ vertex and $\rho K
\bar{K}$ vertex. Similarly, we define
 \bea F_3 &\equiv& \left\{ \begin{array}{llc}
 F(\Lambda_{\psi \rho \pi}=4.5) F(\Lambda_{\rho
 \pi\pi}=1.5)&, & J/\psi \stackrel{\rho}{\to} \pi \pi \pi \\
 F(\Lambda_{\psi \rho \pi}=4.5) F(\Lambda_{\omega \rho \pi}=1.5)&, & J/\psi \stackrel{\rho}{\to} \omega \pi \pi
 \\
 F(\Lambda_{\psi \rho \pi}=4.5) F(\Lambda_{\rho K \bar{K}}=1.5-4.5)&, & J/\psi \stackrel{\rho}{\to} K \bar{K}\pi
 \end{array} \right.
 \eea
\bea F_4 &\equiv& \left\{ \begin{array}{llc}
 F(\Lambda_{\psi \rho \pi}=1.5) F(\Lambda_{\rho
 \pi\pi}=1.5)&, & J/\psi \stackrel{\rho}{\to} \pi \pi \pi \\
 F(\Lambda_{\psi \rho \pi}=1.5) F(\Lambda_{\omega \rho \pi}=1.5)&, & J/\psi \stackrel{\rho}{\to} \omega \pi \pi
 \\
 F(\Lambda_{\psi \rho \pi}=1.5) F(\Lambda_{\rho K \bar{K}}=1.5-4.5)&, & J/\psi \stackrel{\rho}{\to} K \bar{K}\pi
 \end{array} \right.
 \eea


The differential decay widths can be evaluated as
\begin{eqnarray}
d\Gamma(J/\psi \to \omega \pi^+ \pi^- )&=&
\frac{(2\pi)^4}{2M_{\psi}}{|{{T_1}}|}^2  d\Phi_3 (p_0; p_1, p_2,
p_3),  \label{gamma1} \\
d\Gamma(J/\psi \to K \bar{K} \pi )&=&
\frac{({2\pi})^4}{2M_{\psi}}({|{{T_{2a}}}|}^2 + {|{{T_{2b}}}|}^2 +
{|{{T_{2c}}}|}^2 + {|{{T_{2d}}}|}^2) \nonumber \\ && d\Phi_3 (p_0;
p_1, p_2, p_3), \label{gamma3}
\end{eqnarray}
with $M_{\psi}$ being the mass of $J/\psi$ and the three body
phase space factor
\begin{equation}
d\Phi_3 (p_0; p_1, p_2,p_3) = \delta^4 (p_0-p_1-p_2-p_3) \frac{d^3
p_1}{{(2 \pi)}^3 2E_1} \frac{d^3 p_2}{{(2 \pi)}^3 2E_2} \frac{d^3
p_3}{{(2 \pi)}^3 2E_3}.
\end{equation}

>From above equations and following the covariant tensor amplitude
method described in detail in Refs.\cite{chung:93 98}, the entire
calculations are straightforward although tedious. The numerical
results are given in the following section.

\section{Numerical results and discussions}  

In the model described so far, there are four relevant coupling
constants, $g_{\psi\rho\pi}$, $g_{\omega\rho\pi}$ and $g_{\rho\pi\pi}$
 (see Eqs.(\ref{psi rho pi}-\ref{psi rho k k})). The $g_{\rho\pi\pi}$
 can be obtained by evaluating the
process $\rho \to \pi\pi $ with various form factors. In an
analogous way, we can obtain $g_{\psi\rho\pi}$ through the
sequential decays $\psi \to \rho\pi$ with $\rho \to \pi \pi$ and
$g_{\omega\rho\pi}$ through  $\omega \to \rho\pi$ with $\rho \to
\pi \pi$ by using the similar approach described in the section
above.  The experimental data of $\rho \to \pi\pi $,
$\psi \to \rho \pi$, $\omega \to \pi \pi \pi$ decay widths are
from the PDG \cite{PDG}. The results are shown in Table
\ref{table:result}.

We use two forms of $\Gamma_\rho$ in Eqs.(\ref{T1},\ref{T2}).
First we take it  as a constant of 0.149 GeV. Then we take  an
energy dependent width including $\pi\pi$, $\omega\pi$ and $K
\bar{K}$ channels

 \be \Gamma_{\rho}(s) =  \Gamma_{\rho\rightarrow\pi\pi}(s)
 +\Gamma_{\rho\rightarrow \omega\pi}(s) + \Gamma_{\rho\rightarrow K
 \bar{K}}(s) \equiv \Gamma_{\rho\rightarrow (\pi\pi+\omega\pi+ K
 \bar{K})},
 \ee
where energy dependent partial width
$\Gamma_{\rho\rightarrow\pi\pi}$, $\Gamma_{\rho\rightarrow
\omega\pi}$ and $\Gamma_{\rho\rightarrow K \bar{K}}$ are obtained
by the method similar to Eqs.(\ref{gamma1},\ref{gamma3}). Usually
only $\rho\to\pi\pi$ width is considered in the $\rho$ propagator.
But in reality, when the invariant mass of the off-peak $\rho$
meson goes above $\omega\pi$ and $K\bar K$ thresholds, the
off-peak $\rho$ can also decay into these new channels and the
corresponding partial decay widths should be included in its total
energy dependent width. This will make the tail of $\rho$
propagator drops faster for energies above the new thresholds than
the usual propagator with constant width. We find that the two
forms of $\Gamma_{\rho}$ have little influence on coupling
constants. So we use the same coupling constants in amplitudes
with the same form factor (see Table \ref{table:result}).

In order to demonstrate how large effect the s-channel
subthreshold $\rho$ exchange may have upon various channels, their
contributions are calculated and compared with experimental decay
widths of the corresponding final states. The results are listed
in Table \ref{table:result}. The $R^\rho_{\omega\pi\pi}$ and
$R^\rho_{K\bar K\pi}$ represent the ratio of theoretical
contribution from the subthreshold $\rho$ exchange to the
experimental width taken from PDG \cite{PDG} for the channels
$\omega\pi^+\pi^-$ and $K\bar K\pi$, respectively:
  \bea
R^\rho_{\omega\pi\pi}&\equiv&{\frac{\Gamma_{th}(J/\psi\stackrel{\rho}{\rightarrow}\omega\pi^+\pi^-)}
{\Gamma_{ex}(J/\psi\rightarrow\omega\pi^+\pi^-)}},\\
 R^\rho_{K\bar K\pi}&\equiv& {\frac{\Gamma_{th}(J/\psi\stackrel{\rho}{\rightarrow} K\bar{K}\pi)}
 {\Gamma_{ex}(J/\psi\rightarrow K\bar{K}\pi)}}.
  \eea
The range of variations for $R^\rho_{K\bar K\pi}$ in Table
\ref{table:result} for form factor $F_3$ and $F_4$ comes from the
variation of $\Lambda$ between 1.5 and 4.5 GeV for the $\rho K
\bar{K}$ vertex.

\begin{center}
\begin{table}
\caption{The coupling constants and the percentages of the
s-channel subthreshold $\rho$ exchange contribution as functions of
the form factors (F. F.)  and of the choice of $\rho$ width.}
\begin{tabular}{|c|c|c|c|c|c|c|}

 \hline
 F. F.&
 $\Gamma_{\rho}$ &
 $g_{\rho\pi\pi}$&$g_{{\scriptscriptstyle J/{\Psi}} \rho
\pi}$&$g_{\omega\rho\pi}$

 &$R^\rho_{\omega\pi\pi}$
 &$R^\rho_{K\bar K\pi}$
\\
&${\scriptstyle(GeV)}$&&${\scriptstyle(GeV^{-1})}$&${\scriptstyle(GeV^{-1})}$&${\scriptstyle(\%)}$&${\scriptstyle(\%)}$

\\ \hline
 $F_1$
 &0.149&8.18&8.61&14.56&21.0 &4.5
 \\ \cline{2-2}\cline{6-7}
$$
 &${\scriptstyle\Gamma_{\rho\rightarrow(\pi\pi+\omega\pi+ K \bar{K})}}$
 &&$\times 10^{-3}$ & &18.3 &4.3

  \\ \hline
 $F_2$
 &0.149&10.69&8.70&14.68&11.1 &4.2
 \\ \cline{2-2}\cline{6-7}
$$
 &${\scriptstyle\Gamma_{\rho\rightarrow(\pi\pi+\omega\pi+ K \bar{K})}}$
 &&$\times 10^{-3}$ &&9.3 &4.0
  \\ \hline

 $F_3$&0.149&6.05&2.31&11.82&7.7 &1.7-3.4
 \\ \cline{2-2}\cline{6-7}

 &${\scriptstyle\Gamma_{\rho\rightarrow(\pi\pi+\omega\pi+ K \bar{K})}}$
 &&$\times 10^{-3}$ &&6.3 &1.6-3.2
  \\ \hline
  $F_4$&0.149&6.05&2.35&11.82&3.9 &1.0-1.8
 \\ \cline{2-2}\cline{6-7}

 &${\scriptstyle\Gamma_{\rho\rightarrow(\pi\pi+\omega\pi+ K \bar{K})}}$
 &&$\times 10^{-3}$ &&3.1 &0.9-1.7
  \\ \hline
\end{tabular}
\label{table:result}
\end{table}
\end{center}

 From Table \ref{table:result}, we see that no
matter which form of $\Gamma_\rho$ and form factor are employed,
the range of the $R^\rho_{\omega\pi\pi}$ is from 3.1\% to 21.0\%
and that of $R^\rho_{K\bar K\pi}$ is from 0.9\% to 4.5\%.
 It means that the
contribution of the s-channel subthreshold $\rho$ exchange is not
negligible for both  $J/\psi \to \omega\pi\pi$ and $J/\psi \to
K\bar{K}\pi$ channels. The subthreshold $\rho$ contribution may
have significant influence on the analysis of resonances near
$\omega\pi$ and $K\bar K$ thresholds.

\begin{figure}[htbp] 
\begin{center}
\includegraphics[scale=0.7]{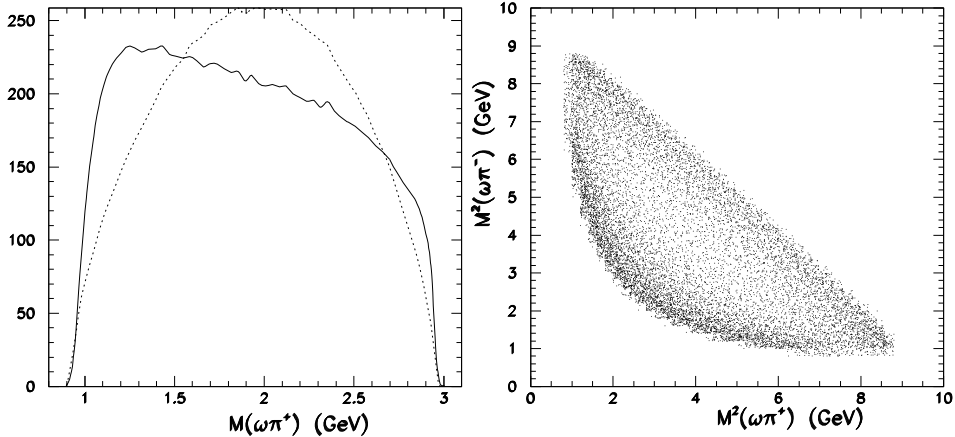}
\caption{ The $\omega\pi$ invariant-mass distribution (solid line)
and the Dalitz plot for $J/\psi \to \omega\pi^+\pi^-$ decay
through $\rho$ exchange with form factor $F_2$ and
$\Gamma_\rho$=0.149 GeV, compared with phase space distribution
(dotted line). }\label{fig:dalitz.omega}
\end{center}
\end{figure}

\begin{figure}[htbp]
\begin{center}
\includegraphics[scale=0.4]{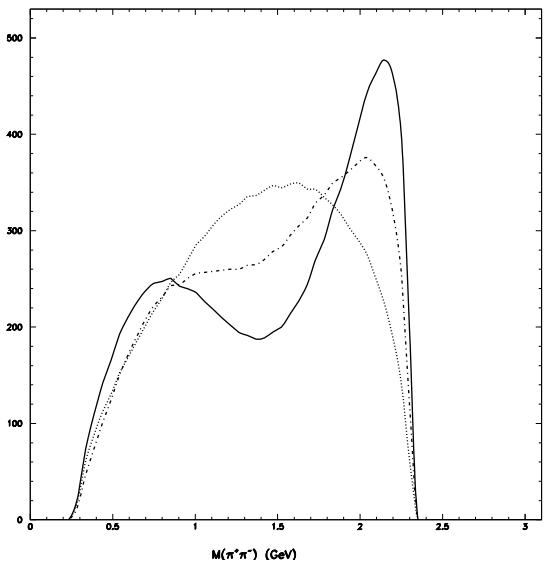}
\caption{The $\pi\pi$ invariant-mass distribution for $J/\psi \to
\omega\pi^+\pi^-$ decay through $\rho$ exchange with form factor
$F_2$ and $\Gamma_\rho$=0.149 GeV (solid line), compared with the
phase space distribution (dotted line) and the result ignoring the
interference effect between subthreshold $\rho^+$ and $\rho^-$
contributions (dot-dashed line). }\label{fig:mass pipi}
\end{center}
\end{figure}

\begin{figure}[htbp] 
\begin{center}
\includegraphics[scale=0.7]{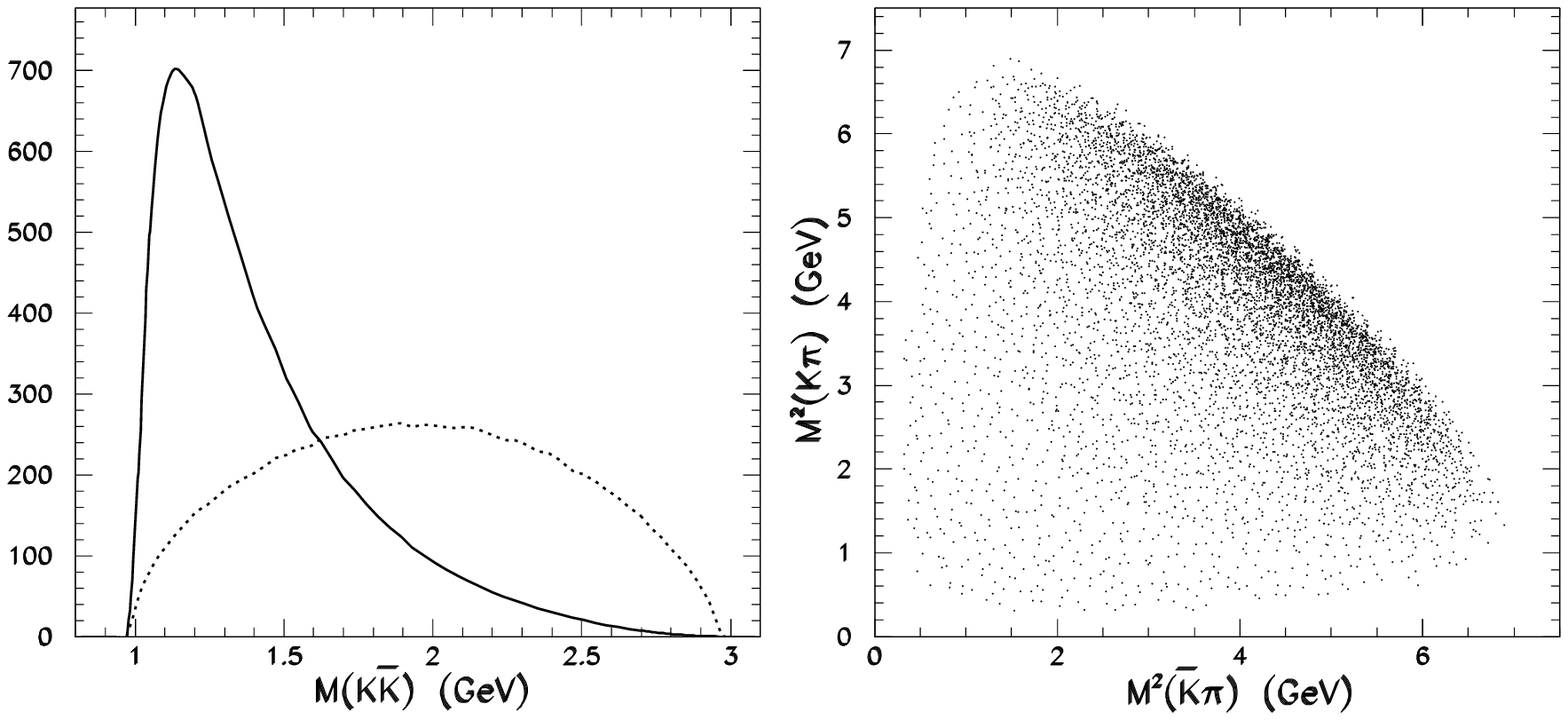}
\caption{The $K \bar{K}$ invariant-mass distribution (solid line)
and the Dalitz plot for $J/\psi \to K\bar{K}\pi$ decay through
$\rho$ exchange with form factor $F_2$ and $\Gamma_\rho$=0.149
GeV, compared with phase space distribution (dotted line).}
\label{fig:invariant.kk}
\end{center}
\end{figure}

In order to see the influence it may have on analysis of resonances
near thresholds,  we perform a Monte Carlo simulation to give
predictions on various invariant mass spectra and Dalitz plots for
these two channels as shown in Fig.\ref{fig:dalitz.omega},
Fig.\ref{fig:mass pipi} and Fig.\ref{fig:invariant.kk}, in which we
take form factor $F_2$ and $\Gamma_\rho$=0.149 GeV. The dotted
lines in  the invariant mass spectra denote the uniform phase space
distributions without considering the dynamical interactions. In
Fig.\ref{fig:dalitz.omega}, a clear enhancement about 1.2 GeV near
the $\omega\pi$ threshold appears in both invariant mass spectrum
and Dalitz plot. Comparing with experimental results (Fig.7 and
Fig.10 in Ref.\cite{dm2 5pi}; Fig.1 and Fig.2 in
Ref.\cite{bes-sigma}), one can expect that this enhancement, as the
background of $b_1(1235)$, should reduce the measured width of
$b_1(1235)$ from this reaction and may well explain the ``$b_1$
puzzle".

One interesting phenomena in the Dalitz plot of
Fig.\ref{fig:dalitz.omega} is that there is also a clear
enhancement band near $\pi\pi$ threshold, which is shown more
clearly in the Fig.\ref{fig:mass pipi} for the $\pi\pi$
invariant-mass distribution of $J/\psi \to \omega\pi^+\pi^-$ decay
(solid line). This enhancement band comes mainly from the
interference terms of two Feynman diagrams in Fig.\ref{fig:omega},
because if we do not include the interference term of Fig.
\ref{fig:omega}, the visible bump located in $0.5-1.0$ GeV  will
be much reduced in the $\pi^+ \pi^-$ invariant-mass distribution
as shown by the dot-dashed line. Therefore, the subthreshold
$\rho$ contribution may even have influence upon the analysis of
the $\sigma$ meson from this channel.

In Fig.\ref{fig:invariant.kk}, a clear peak around 1.1 GeV near
the threshold is also seen in the $K \bar{K}$ invariant-mass
distribution.  By looking at Table \ref{table:result} and
Fig.\ref{fig:invariant.kk},  it is natural to expect that the
subthreshold $\rho$ contribution is an important source for the
near-threshold enhancement found in $J/\psi \to K^+ K^- \pi^0$
decay by the DM2 Collaboration \cite{dm2-KKpi}. Hence the result
provides another evidence of the important role played by the
subthreshold $\rho$ contribution.

Since the mass of $\rho$ is more than 150 MeV below the
$\omega\pi$ and $K \bar{K}$ thresholds and the width of  $\rho$ is
not very broad, some people naively assume that the  subthreshold
$\rho$ contribution can be neglected in these channels of $J/\psi$
decays.  However, this paper should greatly change this point of
view for the channels with final state particles coupling strongly
to the subthreshold $\rho$. A similar important subthreshold
contribution was previously noticed for $J/\psi\to\bar NN\pi$
channels from subthreshold nucleon pole \cite{Sinha}.

In summary, the $\rho$ exchange plays a very important role in
many low-energy strong interaction processes, such as $\pi\pi$
scattering, $\pi  K$ scattering, $\pi N$ interaction etc..   In
this paper, we extend the mechanism to interpret some long
standing problems observed in $J/\psi$ decays. It is found that
the subthreshold $\rho$ contribution is not negligible for both
$J/\psi\to\omega\pi\pi$ and $J/\psi\to K\bar K\pi$ channels, and
should be included in analyzing these channels. It may well
explain the longstanding $b_1$ puzzle near $\omega\pi$ threshold
in $J/\psi\to\omega\pi\pi$ and the near $K\bar K$ threshold
enhancement in $J/\psi \to K \bar{K} \pi$ decay.

\section*{Acknowledgments}
We acknowledge stimulating discussions with BES members on
relevant issues. One of us (Wu) would like to thank Bo-Chao Liu,
Ju-Jun Xie, Xin Zhang, Feng-Kun Guo, Hai-Qing Zhou for valuable
comments and discussions during the preparation of this paper.
This work is partly supported by the National Nature Science
Foundation of China under grants Nos. 10225525, 10435080 and by
the Chinese Academy of Sciences under project No. KJCX2-SW-N02.


\end{document}